\documentclass[aps,twocolumn,floatfix,superscriptaddress,preprintnumbers,showpacs,showkeys]{revtex4}
\usepackage{slashed}
\usepackage{amssymb}
\usepackage{amsmath}
\usepackage{amsfonts}
\usepackage{epsfig}
\usepackage{graphicx}
\usepackage{tabularx}
\newcommand{\seq}{\begin{subequations}}
\newcommand{\sen}{\end{subequations}}
\newcommand{\eq}{\begin{eqnarray}}
\newcommand{\en}{\end{eqnarray}}

\def\shiftdown#1{#1\llap{\lower.04ex\hbox{#1}}}

\newcommand{\ra}{\rangle}
\newcommand{\la}{\langle}

\newcommand{\bfk}{{\bf k}_{\perp}}

\def\arraystretch{1.5}

\begin{document}

\title{Nonperturbative contribution to the strange-antistrange asymmetry  
of the nucleon sea} 

\author{Alfredo Vega}
\affiliation{Instituto de F\'isica y Astronom\'ia y Centro 
de Astrof\'isica de Valpara\'iso, 
Universidad de Valpara\'iso,\\
A. Gran Breta\~na 1111, Valpara\'iso, Chile} 

\author{Ivan Schmidt}
\affiliation{Departamento de F\'\i sica y Centro Cient\'\i
fico Tecnol\'ogico de Valpara\'\i so (CCTVal), 
Universidad T\'ecnica Federico Santa Mar\'\i a, 
Casilla 110-V, Valpara\'\i so, Chile} 

\author{Thomas Gutsche}
\affiliation{Institut f\"ur Theoretische Physik, 
Universit\"at T\"ubingen,
Kepler Center for Astro and Particle Physics,
Auf der Morgenstelle 14, D-72076, T\"ubingen, Germany}

\author{Valery E. Lyubovitskij}
\affiliation{Institut f\"ur Theoretische Physik, 
Universit\"at T\"ubingen,
Kepler Center for Astro and Particle Physics,
Auf der Morgenstelle 14, D-72076, T\"ubingen, Germany}
\affiliation{
Department of Physics, Tomsk State University,
634050 Tomsk, Russia}
\affiliation{Mathematical Physics Department,
Tomsk Polytechnic University,
Lenin Avenue 30, 634050 Tomsk, Russia}

\begin{abstract}

We give predictions for the nonperturbative (intrinsic) contribution 
to the $s(x) - \bar{s}(x)$ asymmetry of the nucleon sea. 
For this purpose we use different light-front wave functions
inspired by the AdS/QCD formalism, together with a model of 
the nucleon in terms of meson-baryon fluctuations. 
The holographic wave functions for an arbitrary number of constituents,  
recently derived by us, give results quite close to known parametrizations 
that appear in the literature. 

\end{abstract}

\date{\today}



\maketitle

\section{Introduction}

Nowadays it is confirmed that hadrons are built by valence quarks together
with a sea of $q \bar{q}$ pairs and gluons. 
This sea plays an important role in the understanding of several hadronic 
properties and 
provides an explanation of many experimental results in hadronic physics 
(for overviews and discussions of experimental and theoretical progress in this 
field see e.g. Refs.~\cite{Sullivan:1971kd}-\cite{Feng:2012gu}).  
In this vein one of the interesting aspects is the study of the strange 
quark sea and the $s(x) - \bar s(x)$ asymmetry, since
both experimental and theoretical analyses indicate the existence 
of an strange-quark asymmetry (SQA) in the nucleon sea. 
Therefore, this asymmetry is a major hadronic observable, whose study 
can shed light on our understanding of the nucleon structure.

There are two main mechanisms that can
generate a SQA in the nucleon sea -- nonperturbative (intrinsic)  and 
perturbative (extrinsic) 
(see e.g. discussion in Refs.~\cite{Catani:2004nc,Feng:2012gu}). 
The nonperturbative (or intrinsic) contribution to the strange-quark asymmetry 
can be considered to originate from nucleons fluctuating into virtual baryon-meson states 
($\Lambda K$ and $\Sigma K$).  
These contributions can be estimated by using nonperturbative models 
for the nucleons~\cite{Signal:1987gz,Brodsky:1996hc,Cao:1999da} and  
are shown to be dominant in the large-$x$ region. 
As was shown in Ref.~\cite{Catani:2004nc}, the perturbative SQA, 
which is significant in the small-$x$ region, 
is produced by perturbative QCD evolution at 
next-to-next-to-leading order (NNLO) or at three loops.  
Such phenomenon occurs even if the SQA vanishes at the 
initial scale due to nonvanishing $u$ and $d$ valence quark 
densities~\cite{Catani:2004nc}. 

In this work we give predictions for the nonperturbative contribution 
to the $s(x)-\bar{s}(x)$ asymmetry, following the light-front approach
proposed by Brodsky and Ma~\cite{Brodsky:1996hc}.  
This approach deals with two-body light-front wave functions (LFWF) 
describing meson-baryon fluctuations of the proton in convolution with 
specific quark LFWFs. 
The result was that $s(x) < \bar{s}(x)$ 
at small-$x$ and $s(x) > \bar{s}(x)$ at large-$x$, which
is a behavior opposite to the one 
obtained in meson cloud models~\cite{Cao:1999da}.  
In Ref.~\cite{Brodsky:1996hc} both
Gaussian and power-law quark LFWFs were considered,  
leading to similar results.  
We intend to clarify this issue,
using what we consider are more realistic LFWFs.
For this purpose we use three different kinds 
of wave functions and show that the asymmetry 
is quite sensitive to the LFWFs. 
Therefore, the study of the $s-\bar s$ asymmetry could serve 
as a further tool to distinguish between these LFWFs. 
In particular, we consider the traditional Gaussian wave 
function~\cite{Brodsky:1996hc} 
and two variants of LFWFs motivated by AdS/QCD models. 
These last ones are extracted by  matching electromagnetic 
form factors in LF QCD and AdS/QCD for the massless 
case, and introducing modifications in order to 
take into account finite quark masses~\cite{Brodsky:2006uqa}. 
Notice that the same matching procedure allows for an extraction of 
GPDs~\cite{Vega:2010ns} and, in addition, for a 
generalization of the LFWFs for hadrons with an arbitrary number 
of constituents~\cite{Gutsche:2013zia}. 
In particular, we consider an holographic LFWF (variant I) obtained 
from matching to LF QCD at large $x$ and a
holographic LFWF (variant II) obtained from matching to LF QCD 
at all values of $x$ and for an arbitrary number of 
constituents~\cite{Gutsche:2013zia}.  
The second type of holographic LFWF is more useful for our purposes, 
because it can be applied to the description 
of hadrons with an arbitrary number of constituents.

We should stress that the all three LFWFs 
can reproduce hadronic experimental data (e.g see \cite{Brodsky:2006uqa,Liu:2015jna}), and therefore the study of the s-sbar asymmetry 
serves as an independent tool to asses which of these functions provides a better 
phenomenological description of the nucleon. 
All considered LFWFs contain the scale parameter 
$\kappa$, which has been fixed at the value that gives the best description of the data.
For completeness we also consider the sensitivity of the $s(x)-\bar{s}(x)$ asymmetry 
to a choice of this parameter.

The paper is structured as follows. In Sec.~II we introduce the main 
ingredients of the light-front approach that we use to calculate the 
$s(x) - \bar{s}(x)$ asymmetry. Then we briefly describe the set of LFWFs 
used in our calculations. In Sec.~III we discuss our results. 
Finally in Sec.~IV we present our conclusions.  

\section{$s(x) - \bar{s}(x)$ asymmetry in a light cone approach}

In the light-front formalism the proton state can be expanded in 
a series of components as 
\eq
| P \rangle &=& | uud \rangle \psi_{uud/p} + |uudg \rangle \psi_{uudg/p} 
\nonumber\\
&+& \sum_{q \bar{q}} |uudq\bar{q} \rangle \psi_{uudq\bar{q}/p} + \ldots 
\en
where $|uud \rangle$, $|uudg \rangle$, $|uudq\bar{q} \rangle$, $\ldots$ 
are the contributing Fock states and $\psi_{uud/p}$, $\psi_{uudg/p}$, 
$\psi_{uudq\bar{q}/p}$, $\ldots$ are the quark/gluon LFWFs corresponding 
to these states. 
In Ref.~\cite{Brodsky:1996hc} a different light front approach was 
proposed, in which the nucleon has components arising from meson-baryon 
fluctuations, while these hadronic components are composite systems 
of quarks. This approach is similar to expansions used
in meson-cloud models~\cite{Sullivan:1971kd}. 
In this case the nonperturbative contributions to the $s(x)$ and 
$\bar{s}(x)$ distributions in the proton can be expressed as convolutions 
(see e.g. Ref.~\cite{Cao:1999da})
\eq
\label{s}
s(x) = \int_{x}^{1} \frac{dy}{y} f_{\Lambda/K \Lambda} (y) q_{s/\Lambda} 
\biggl( \frac{x}{y} \biggl)\,,\\
\label{sbar}
\bar{s}(x) = \int_{x}^{1} \frac{dy}{y} f_{K/K \Lambda} (y) q_{\bar{s}/K} 
\biggl( \frac{x}{y} \biggl)\,,
\en
where $q_{s/\Lambda}$ and $q_{\bar{s}/K}$ are distributions of $s$ quarks 
and $\bar{s}$ antiquarks in the $\Lambda^0 (\Sigma^0)$ and $K^{+}$, 
respectively.  
The functions $f_{\Lambda/K \Lambda}(y)$ and $f_{K/K\Lambda}(y)$ describe 
the probability to find a $\Lambda$ or a $K$ with light-front momentum 
fraction y in the $K\Lambda$ state.

In the light-front model proposed in~\cite{Brodsky:1996hc} the meson-baryon 
distribution functions are calculated through the relation 
\eq 
\label{Fluctuacion}
f_{B/BM}(y) = \int \frac{d^2\bfk}{16 \pi^3} |\psi_{BM}(y,\bfk)|^{2}.
\en
where $\psi_{BM}(y,\bfk)$ is the LFWF describing the distribution of the 
baryon-meson (BM) components. 
An important property of these functions, 
which follows from momentum and charge conservation, 
is that $f_{BK}(y) = f_{KB}(1-y)$~\cite{Holtmann:1996be}.  
Here we have defined
$f_{BK}(y) = f_{B/BK}(y)$ and $f_{KB}(y) = f_{K/BK}(y)$.
In this work we consider a fluctuation probability  for 
$N \rightarrow \Lambda K$ of $1.27 \%$,  
as in Ref.~\cite{Cao:1999da}. 

In the same manner, the distribution functions $q_{s/\Lambda}$ and 
$q_{\bar{s}/K}$ can be determined by
\eq 
\label{sLambda}
q_{s/\Lambda}(x) &=& \int\frac{d^2\bfk}{16 \pi^3} |\psi_{\Lambda}(x,\bfk)|^{2} 
\,,\label{sK}\\
q_{\bar{s}/K}(x) &=& \int \frac{d^2\bfk}{16 \pi^3} |\psi_{K}(x,\bfk)|^{2}\,.
\en
To calculate $f_{\Lambda/K \Lambda}$, $f_{K/K \Lambda}$, 
$q_{s/\Lambda}$ and $q_{\bar{s}/K}$ it is necessary to know the LFWF 
for the distribution of 
$\Lambda$ and $K$ inside the states $\Lambda K$ and $\Sigma K$, 
and for the quarks/antiquarks in the $\Lambda$ and $K$. 
As we mentioned before, in this paper we use three different kinds 
of the LFWFs: i) a typical Gaussian LFWF~\cite{Brodsky:1996hc}, ii) a 
so-called holographic LFWFs obtained using light-front holography 
ideas~\cite{Brodsky:2006uqa} 
at large $x$ and iii) a further generalization which is extracted 
from matching at any value of $x$ and that further allows to describe 
hadrons with an arbitrary number of constituents~\cite{Gutsche:2013zia}. 
In the first two cases we directly follow the ideas of 
Ref.~\cite{Brodsky:1996hc}, where two-body wave functions $\psi_{BM}$ 
are formed by two clusters - baryon (as a quark-diquark bound state) and 
meson (as the usual quark-antiquark bound state). 
The third LFWF considers hadrons with an arbitrary number of constituents 
in the two-body approximation. 
In particular, the twist-5 wave function corresponds to the LFWF of 
the baryon-meson bound state $\psi_{BM}$. 
For the inclusion of massive constituents (baryon and meson) 
we follow the procedure 
proposed and realized in Refs.~\cite{Brodsky:2006uqa}.  
In the next subsections we give details of all LFWF used in our calculations. 

\subsection{Gaussian wave function}

Brodsky and Ma~\cite{Brodsky:1996hc} 
suggested to use two-body Gaussian and power-law 
wave functions to calculate the $s(x)$ and $\bar{s}(x)$ asymmetry. 
They got similar results in both cases. 
Here we first consider the Gaussian wave function specified as 
\eq
\psi(x,\bfk) = A \exp\biggl[ - \frac{1}{8 \kappa^2} 
\biggl( \frac{\bfk^2}{x (1-x)} + \mu_{12}^2 \biggr) \biggr] 
\,, 
\en
where 
\eq 
\mu_{12}^{2} = \frac{m_1^2}{x} + \frac{m_2^2}{1-x} 
\en
and 
$m_1$ and $m_2$ are the masses of the constituents. 

In this approach, the functions $f_{\Lambda/K\Lambda}$ and $f_{K/K\Lambda}$ 
are calculated as
\eq 
f_{\Lambda/K\Lambda}(x) 
&=& \frac{\kappa^2 A_{MB}^2}{4 \pi^2} \, x (1-x) \, 
\exp\biggl[ - \frac{\mu_{\Lambda K}^2}{4 \kappa^{2}} \biggr] 
\,, \\
f_{K/K\Lambda}(x) 
&=& \frac{\kappa^2 A_{MB}^2}{4 \pi^2} \, x (1-x) \, 
\exp\biggl[ - \frac{\mu_{K \Lambda}^2}{4 \kappa^{2}} \biggr] 
\,. 
\en
These LFWFs satisfy the constraint $f_{BK}(x) = f_{KB}(1-x)$. 
On the other side, the distribution functions $q_{s/\Lambda}$ and 
$q_{\bar{s}/K}$ are given by
\eq 
q_{s/\Lambda}(x) 
&=& \frac{\kappa^2 A_{B}^2}{4 \pi^2} \, x (1-x) \, 
\exp\biggl[ - \frac{\mu_{s D}^2}{4 \alpha^{2}} \biggr] \,,\\
q_{\bar{s}/K}(x) 
&=& \frac{\kappa^2 A_{M}^2}{4 \pi^2} \, x (1-x) \, 
\exp\biggl[ - \frac{\mu_{\bar{s} K}^2}{4 \kappa^{2}} 
\biggr] \,. 
\en

\begin{center}
\begin{figure}
  \begin{tabular}{c}
        \includegraphics[width=3.5 in]{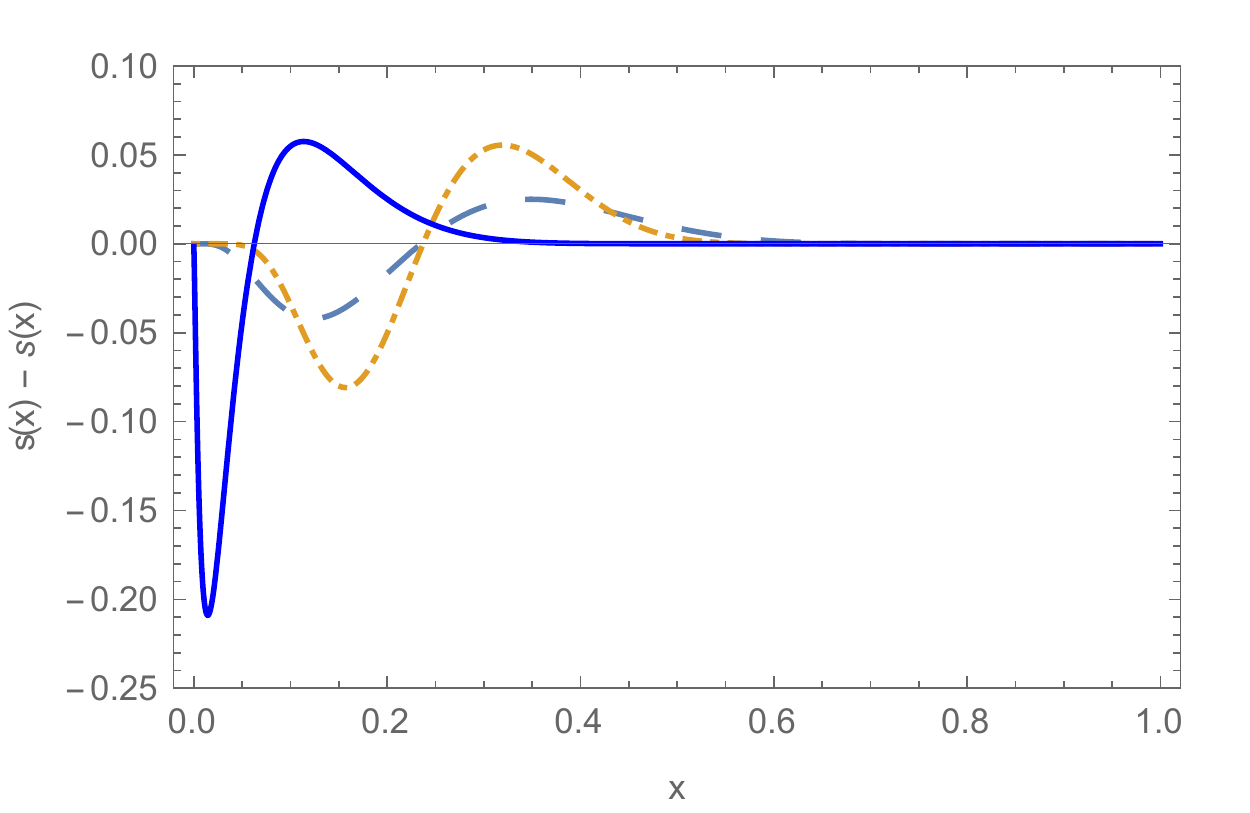}
  \end{tabular}
\caption{$s(x) - \bar{s}(x)$ plots for three different types of LFWFs: Gaussian LFWF (large dashed line -- $\kappa = 330$ MeV), holographic LFWF (variant I), 
(dot dashed line -- $\kappa = 350$ MeV) and holographic LFWF (variant II)(continuous line -- $\kappa = 350$ MeV).}
\end{figure}
\end{center}

Note that all the parameters involved in our calculations --- 
constituent masses of quarks $m_{q} = 330$~MeV, $m_{s} = 480$~MeV, 
of diquark $m_{D} = 600$~MeV, characteristic internal scale 
$\kappa = 330$~MeV --- have been fixed in Ref.~\cite{Brodsky:1996hc} 
and we use exactly the same set of parameters.  
The masses of hadrons involved in the calculation are 
$m_{\Lambda} = 1115.683$~MeV and $m_{K^{+}} = 493.677$~MeV. 
The normalization constants $A_{MB}$, $A_{B}$ and $A_{M}$ are obtained 
considering that the meson-baryon ($f_{\Lambda/K\Lambda}$, $f_{K/K\Lambda}$) 
and quark distribution ($q_{s/\Lambda}$, $q_{\bar{s}/K}$) 
functions are normalized to one: 
\eq 
\hspace*{-.3cm}
\int\limits_0^1 \!dx f_{K(\Lambda)/K\Lambda}(x) \! = \!  
\int\limits_0^1 \!dx q_{s/\Lambda}(x) \! = \! 
\int\limits_0^1 \!dx q_{\bar{s}/K}(x) \! = \! 1 \,. 
\en

Actually, these normalizations are correct 
when we do not include the probability for a meson-baryon
fluctuation. When this probability is included, which is actually
the case for equations~(\ref{s}) and~(\ref{sbar}), then the meson-baryon
distribution normalizations should contain this probability. 
As mentioned before, here we have taken the fluctuation probability for 
$N \rightarrow \Lambda K$ to be $1.27 \%$.

\subsection{Holographic wave functions}

In this section we consider two-body wave functions obtained by using 
light-front holography~\cite{Brodsky:2006uqa}. 
Specifically we use two types of 
wave functions -- variant I (obtained from matching to LF QCD 
at large $x$) and variant II (obtained from matching at all values of $x$ and 
for an arbitrary number of constituents)   
\eq 
\psi(x,\bfk) &=& \frac{A}{\sqrt{x(1-x)}} \nonumber\\
&\times&\exp\biggl[ - \frac{1}{2 \kappa^{2}} 
\biggl( \frac{\bfk^2}{x(1-x)} + \mu_{12}^2 \biggr) \biggr]  
\en
and 
\eq 
\label{PsiHTau}
\hspace*{-.8cm}\psi_{\tau} (x,\bfk) &=& A_{\tau} f_{\tau}(x) \nonumber\\ 
&\times&\exp\biggl[- \frac{x \log(1/x)}{2 \kappa^2 (1-x)} 
\biggl( \frac{\bfk^2}{x(1-x)} + \mu_{12}^2 \biggr) \biggr]\,,
\en   
where 
\eq 
f_{\tau}(x) = \frac{4 \pi}{\kappa} \sqrt{\log(1/x)} 
(1-x)^{\frac{\tau - 4}{2}}\,. 
\en 
Here $\tau$ is the twist of the operator that creates these states, 
which in this case coincides with the number of constituents of the 
particular state.

\begin{center}
\begin{figure}
  \begin{tabular}{c}
        \includegraphics[width=3.0 in]{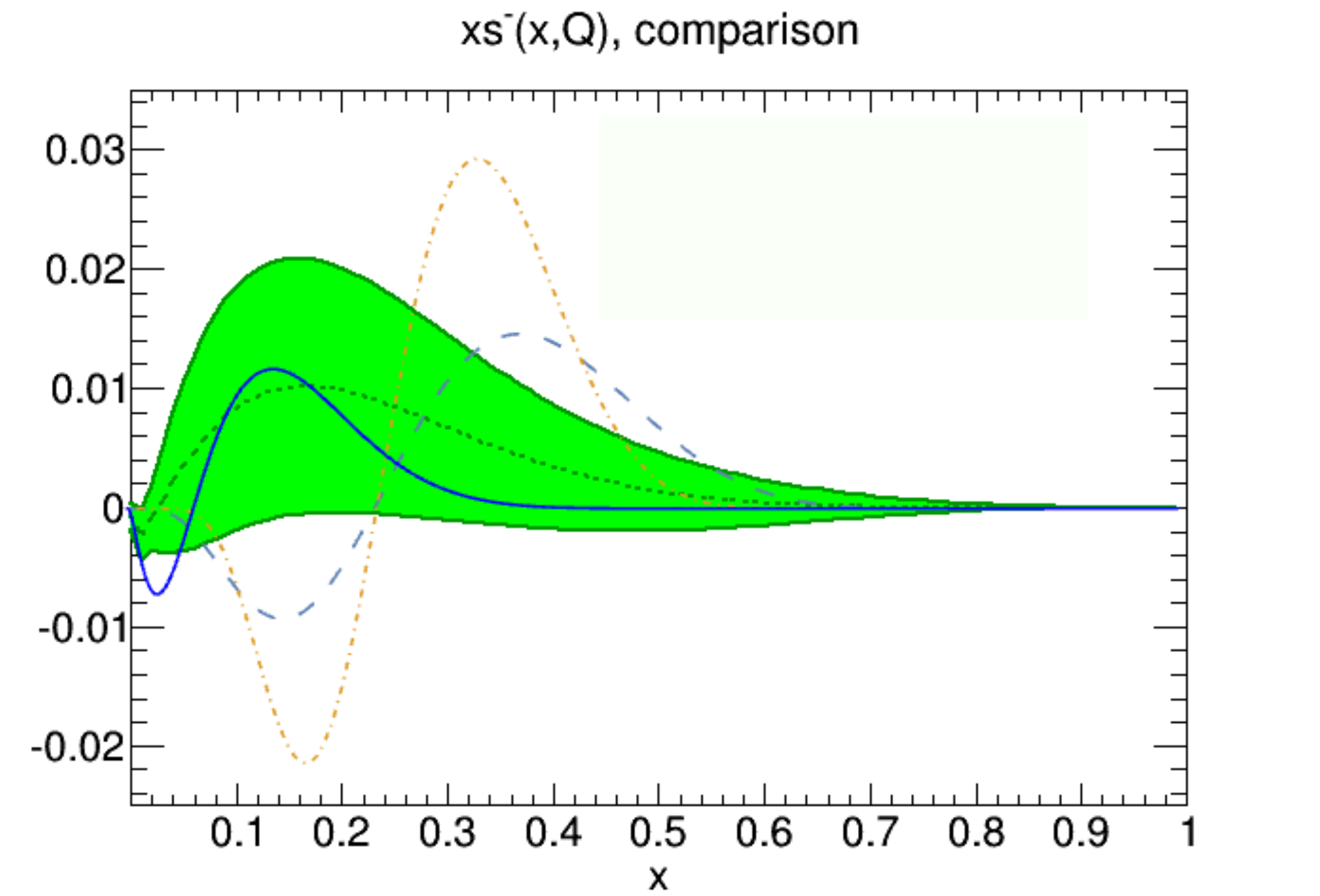}
  \end{tabular}
\caption{$x S^{-} = x(s(x) - \bar{s}(x))$. Green region and small 
dashed line correspond to MMHT\cite{Harland-Lang:2014zoa}) 
that it was generated with APFEL~\cite{Carrazza:2014gfa}. 
Other lines correspond to same cases in Fig. 1.} 
\end{figure}
\end{center}

The first function satisfies the constraint $f_{BM}(x) = f_{MB}(1-x)$ with
masses which are the same as used in the Gaussian case. 
The second LFWF~\cite{Gutsche:2013zia} is also a two-body wave function, 
describing both clusters as quark-diquark bound states. 
For both holographic LFWFs we take the same value $\kappa = 350$ MeV 
fixed previoously from the best description of data on hadron 
properties~\cite{Gutsche:2013zia}. 

In our specific calculations of the meson-baryon distribution functions 
we use $\tau = 5$ with masses $m_{1} = m_{B}$ and $m_{2} = m_{M}$. 
Note that the condition $f_{BM}(x) = f_{MB}(1-x)$ is not satisfied 
in the case of the second holographic LFWF.  
To avoid this problem we define the meson-baryon distribution functions as 
\eq 
\label{Fluctuacion2_1}
f_{BM}(x) &=& \int \frac{d^2\bfk}{16 \pi^3} |\phi_{BM}(x,\bfk)|^{2}\,, 
\nonumber\\
\phi_{BM}(x,\bfk) &=& \psi_{5}(x,\bfk) + \psi_{5}(1 - x,\bfk) \,. 
\en
To calculate $q_{s/\Lambda}$ and $q_{\bar{s}/K}$ 
we directly use equation~(\ref{PsiHTau}) with $\tau = 3$ 
and $2$ respectively. 
The parameters used in this case are the same as 
in the case of the holographic LFWF (variant II), also normalized to one.

\begin{center}
\begin{figure}
  \begin{tabular}{c}
        \includegraphics[width=3.5 in]{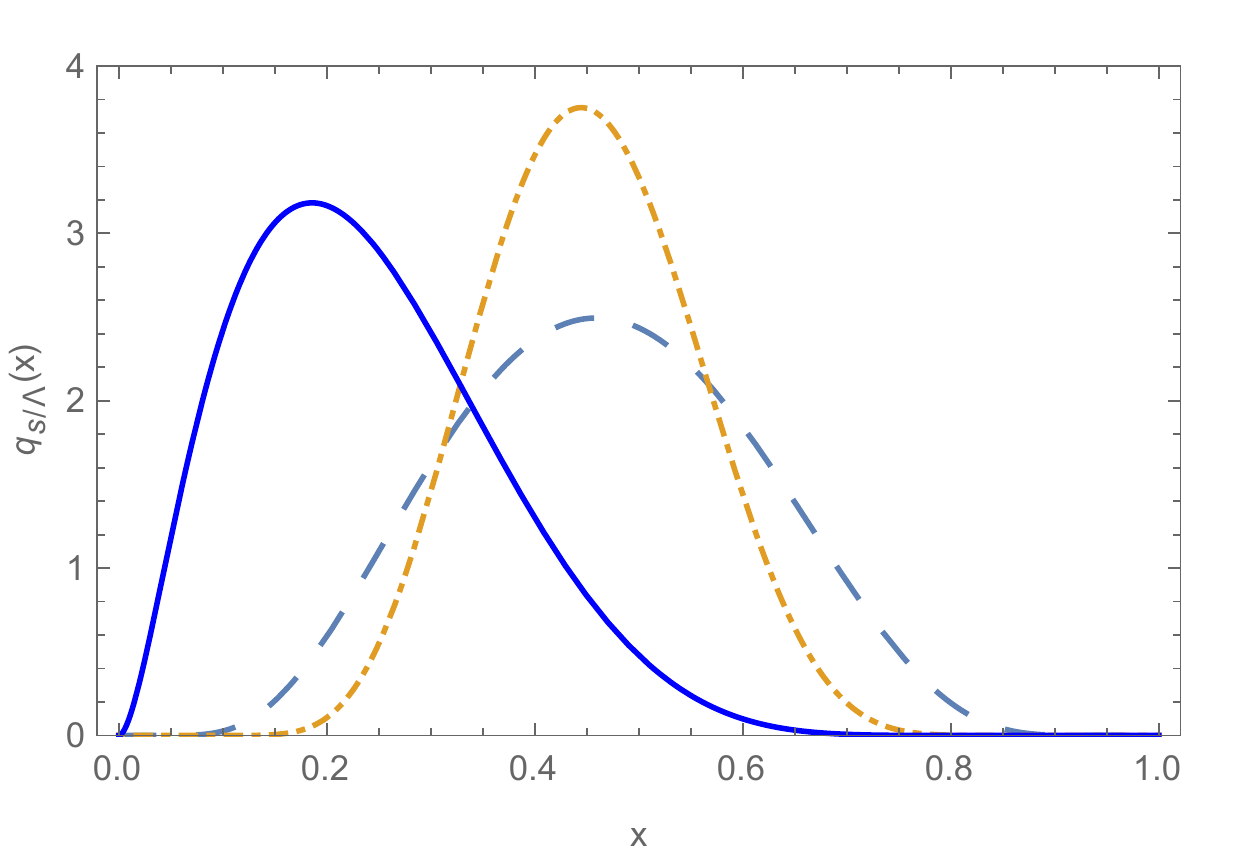}
  \end{tabular}
\caption{Strange quark density $q_{s/\Lambda}$. 
All notations as in Fig.~1.} 
\end{figure}
\end{center}

\section{Results}

\begin{table}[hb]
\begin{center}
\caption{Second moment $\la x (s - \bar{s}) \ra$} 

\vspace*{.1cm}

\def\arraystretch{1.1}
\begin{tabular}{|c|l|l|}
\hline
$\kappa$ (MeV) & \quad\quad Gaussian LFWF \\ 
\hline 
330            & \quad\quad 0.00134       \\
500            & \quad\quad 0.00108       \\
1000           & \quad\quad 0.00058       \\ 
\hline 
$\kappa$ (MeV) & \quad\quad Holographic LFWF (I)  \\
\hline 
350            & \quad\quad 0.00157       \\
550            & \quad\quad 0.00150       \\
700            & \quad\quad 0.00143       \\
\hline 
$\kappa$ (MeV) & \quad\quad Holographic LFWF (II) \\
\hline 
350            & \quad\quad 0.00091       \\
550            & \quad\quad 0.00065       \\
700            & \quad\quad 0.00047       \\
\hline
\end{tabular}
\end{center}
\end{table}

For each LFWF we have calculated the meson-baryon distribution 
functions $f_{\Lambda/K\Lambda}$, $f_{K/K \Lambda}$ and 
the quark distributions $q_{s/\Lambda}$ and $q_{\bar{s}/K}$. 
Then using these quantities we calculate $s(x) - \bar{s}(x)$. 
Fig.~I shows results for all three types of LFWFs using two slightly 
different values for the scale parameter $\kappa$, for reasons explained before.
Notice that in both the Gaussian and the first holographic case 
(variant I) the point where the
asymmetry vanishes is at $x \sim 0.7$, whereas in the other holographic 
case (variant II) it is near $x \sim 0.35$. 
Fig.~2 shows results for $x (s(x) - \bar{s}(x))$ 
calculated with models discussed here. Considering the value and location of maximum in $x(s(x) - \bar{s}(x))$, and region where start to be zero,
we can see that our result for the holographic LFWF (variant II) 
is consistent with the result of the fit done in MMHT~\cite{Harland-Lang:2014zoa}.

\begin{center}
\begin{figure}
  \begin{tabular}{c}
        \includegraphics[width=3.5 in]{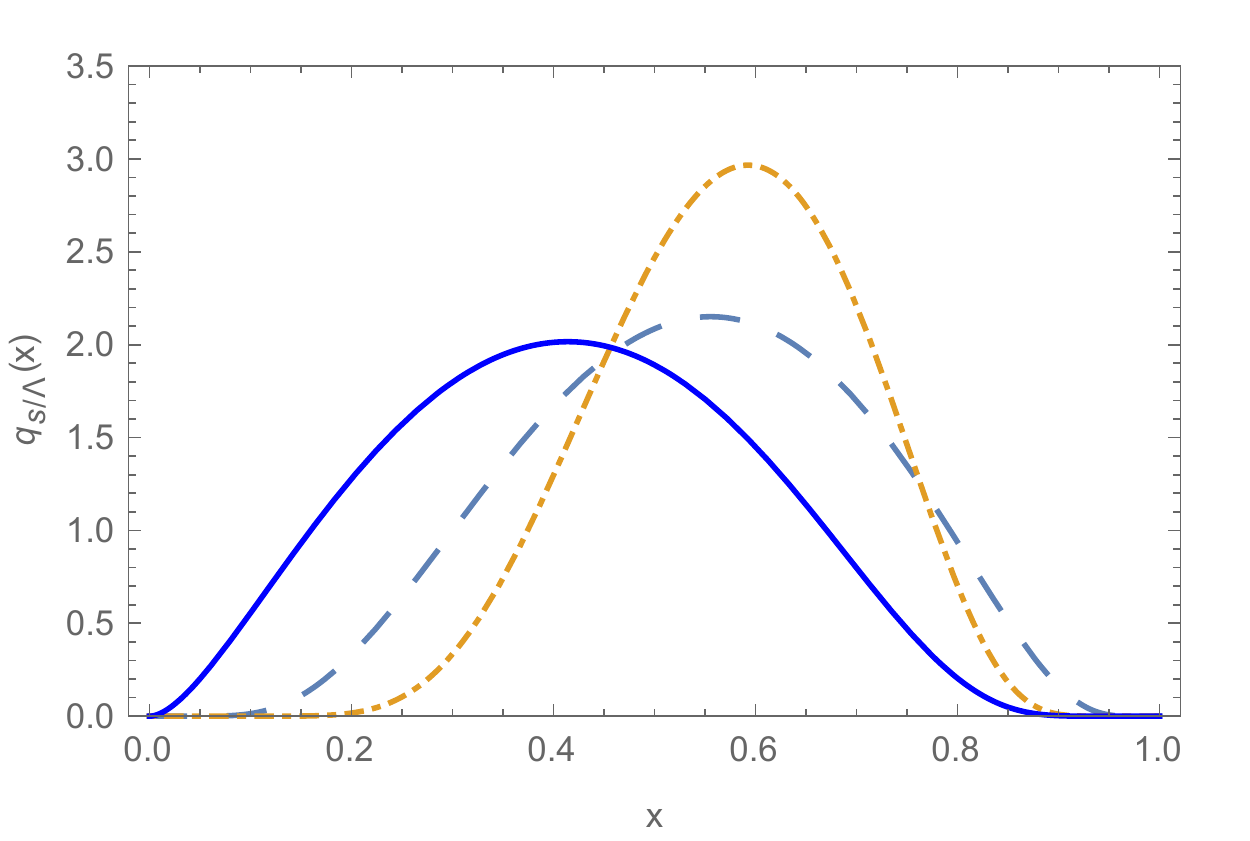}
  \end{tabular}
\caption{Strange quark density $q_{\bar{s}/K^{+}}$.  
All notations as in Fig.~1.} 
\end{figure}
\end{center}

Figs.~3 and 4 show our predictions for the $s$- ($q_{s/\Lambda}$)  
and $\bar s$-quark ($q_{\bar{s}/K^{+}}$) distributions for the three variants 
of the LFWF. By inspection of Figs.~2 and 3 we can conclude that 
the results for the Gaussian and holographic (variant I) LFWF are close 
to each other, while the results for the holographic LFWF (variant II) are different. For the holographic LFWF (variant II) we get consistency with the results of the global fit performed 
by MMHT\cite{Harland-Lang:2014zoa}.

Finally in Table I we present the results for the second moment  
$\la x (s - \bar{s}) \ra$, using the different LFWFs. 
This quantity is defined as 
\eq 
\la x (s - \bar{s}) \ra = \int\limits_0^1 dx x \Big( s(x) - \bar s(x) \Big) 
\,, 
\en 
where the $s(x)$ and $\bar s(x)$ are defined in Eqs.~\ref{s} and ~\ref{sbar}. 
All our results for the second moment are small, positive and 
consistent with a value $\la x (s - \bar{s}) \ra \approx 0.005$ 
mentioned in Ref.~\cite{Catani:2004nc}, which would be required to 
attribute the NuTeV anomaly~\cite{Zeller:2001hh} to the strange 
asymmetry alone. The value is also in a good agreement with 
predictions of different models of $| \la x (s - \bar{s}) \ra | \sim 10^{-4}$ 
(see discussion in Ref.~\cite{Catani:2004nc}). 
For completeness, we also quote some other results for the second moment  
generated in nonperturbative mechanisms. The result 
$\la x (s - \bar{s}) \ra = - 0.0027 \pm 0.0013$~\cite{Zeller:2002du} 
was deduced from a lowest-order QCD analysis of neutrino data, 
$- 0.001 <\la x (s - \bar{s}) \ra < 0.004$ from a 
global QCD fit done in Ref.~\cite{Olness:2003wz}. 
Finally we mention the result for the second moment generated in 
a perturbative mechanism in Ref.~\cite{Catani:2004nc} of  
$\la x (s - \bar{s}) \ra \approx - 5 \times 10^{-4}$ at the scale 
$Q^2 = 20$ GeV$^2$. Last result does not change too much when evolved to 
low scales and lies in the band derived in Ref.~\cite{Olness:2003wz}. 

\section{Conclusions}

We calculated the $s(x) - \bar{s}(x)$ asymmetry 
in a light-front model considering three types of LFWFs. 
In all of these cases we observe that $s(x) < \bar{s}(x)$ for 
small values of $x$ and $s(x) > \bar{s}(x)$ in the region of large $x$.
This behavior is exactly opposite to the one obtained in meson-cloud
models. 
For the cases of the Gaussian 
and the holographic (variant I) LFWFs the asymmetry vanishes at $x \sim 0.6$, 
while for the second variant of the holographic LFWF this point moves to a
lower value of $x \sim 0.35$.
The latter case is closer to
MMHT~\cite{Harland-Lang:2014zoa} (other recent parametrizations are NNPDF~\cite{ Ball:2014uwa}, 
MSTW~\cite{Martin:2009iq}, MMHT~\cite{Harland-Lang:2014zoa} and nCTEQ~\cite{Kusina:2015vfa}). Finally, our 
predictions for the second moment of the strange quark asymmetry 
$\la x (s - \bar{s}) \ra = 0.00102 \pm 0.00055$ are consistent 
with previous results obtained with nonperturbative mechanisms.

Holographic wave functions has been used to calculate hadron properties, and his parameters was fixed, therefore, the study of the s-sbar asymmetry 
serves as independent tool to distinguish these functions. 
We present the results for fixed value of the $\kappa$, although for completeness also allow 
a variation of this parameter to display the sensitivity of s-sbar asymmetry 
to a choice of this parameter.


\begin{acknowledgments}

The authors thank Stan Brodsky and Guy de T\'e\-ra\-mond for useful 
discussions. This work was supported   
by the German Bundesministerium f\"ur Bildung und Forschung (BMBF)
under Project 05P2015 - ALICE at High Rate (BMBF-FSP 202):
``Jet- and fragmentation processes at ALICE and the parton structure  
of nuclei and structure of heavy hadrons'', 
Tomsk State University Competitiveness Improvement Program 
and the Russian Federation program ``Nauka'' (Contract No. 0.1526.2015, 3854), 
by FONDECYT (Chile) under Grant No. 1100287 and Grant No. 1141280 and 
by CONICYT (Chile) Research Project No. 80140097,
and under Grant No. 7912010025. 
V. E. L. would like to thank Departamento de F\'\i sica y Centro
Cient\'\i fico Tecnol\'ogico de Valpara\'\i so (CCTVal), Universidad
T\'ecnica Federico Santa Mar\'\i a, Valpara\'\i so, Chile 
and Instituto de F\'isica y Astronom\'ia, Universidad de Valpara\'iso, Chile
for warm hospitality.

\end{acknowledgments}

\end{document}